# Dispersion behavior of two dimensional monochalcogenides

**Abdus Salam Sarkar[1]\* and Emmanuel Stratakis[1,2]\***

[1]Institute of Electronic Structure and Laser, Foundation for Research and Technology-Hellas, Heraklion, 700 13 Crete, Greece.

[2]Physics Department, University of Crete, Heraklion, 710 03 Crete, Greece.

Email: salam@iesl.forth.gr ; stratak@iesl.forth.gr

**Abstract**

Solution processable two-dimensional (2D) materials have provided an ideal platform for both fundamental studies and wearable electronic applications. Apart from graphene and 2D dichalcogenides, $IV_A$-VI monochalcogenides (MMCs) has emerged recently as a promising candidate for next generation electronic applications. However, the dispersion behavior, which is crucial for the quality, solubility and stability of MMCs, has been quite unexplored. Here, the exfoliation and the dispersion behavior of Germanium (II) monosulfide (GeS) and Tin (II) monosulfide (SnS) nanosheets has been investigated in a wide range of organic solvents. Nine different organic solvents were examined and analyzed, considering the solvent polarity, surface tension, and Hansen solubility parameters. A significant yield of isolated GeS and SnS flakes, namely ~16.4 and ~23.08 μg/ml in 2-propanol and N-Methyl-2-pyrrolidone respectively were attained. The isolated flakes are few-layers nanosheets with lateral sizes over a few hundreds of nanometers. The MMCs colloids exhibit long-term stability, suggesting the MMCs applicability for scalable solution processable printed electronic device applications.

1. **Introduction**

Since the successful isolation of two dimensional (2D) graphene [1], 2D materials have attracted extraordinary attention due to its potential application in next generation electronics, optoelectronics, and photonics [2-5]. In particular, black phosphorene analogous group $IV_A$-VI (Ge/Sn-S) metal monochalcogenides (MMCs) have emerged as rising star materials for fundamental research as well as next generation electronic and photonic applications [3, 6-16], including thin film transistor, solar photovoltaics, sensors, and saturable absorbers [3, 7, 9, 16]. This is attributed to unusual intrinsic in-plane anisotropic physical properties complemented with a high carrier mobility, directional dependent bandgap, odd-even quantum confinement effect and large absorption coefficient of such materials [2, 3, 6, 9, 17, 18]. On top of that, the earth abundant and environmentally friendly behavior of GeS and SnS make them suitable for practical use.

Physical vapor deposition (PVD) and micromechanical exfoliation (ME) are the most widely used synthesis methods of GeS and SnS [6, 8-10, 13, 19]. However, the PVD grown GeS and SnS samples are deposited at high temperature and require a specific substrate [8, 9]. Currently, the large area and thin layer deposition on arbitrary substrates with uniform crystal quality at a low processing temperature is not possible. This is a significant limitation for mass production and roll-to-roll, electronic applications. Besides this, the low exfoliation yield and the absence of dimensional reproducibility in micromechanical exfoliation, harshly restrict their applicability in large area solution processable electronics[3, 19]. Another complexity is the high interlayer binding energy, which leads to a large electron distribution and electronic coupling between the adjacent MMCs layers and therefore to low-probability in mechanical isolation of single layer nanosheets.

The liquid phase exfoliation (LPE) is an ideal alternative for the mass production and large area applications of MMCs, including solution processable or wearable electronics [3, 16, 20, 21]. Using LPE methods, large quantities can be isolated from bulk powders and can be dispersed in various organic solvents. In particular, the LPE- exfoliated Ge/Sn-S exhibited very low dimensionality, i.e. less than 5 nm nanosheet thicknesses in common organic solvents [7, 22-25]. However, the exfoliation and isolation of GeS and SnS colloids in various organic solvents is not a straightforward process, as several critical parameters are involved including the exfoliation yield and the chemical stability. Moreover, the systematic isolation of monolayers or even ultrathin layers is quite challenging [7, 24]. In this context, there is a huge gap in the literature on exfoliation yield, solubility, stability and, in general, the dispersion behavior of isolated colloids in various organic solvents. Moreover, there is a lack of systematic studies on the LPE of GeS and SnS inks in different organic solvents. Therefore, it is timely to evaluate the dispersion behavior of GeS and SnS colloids in various solvents to acquire important information on the compatibility, dispersibility and stability of such MMC inks for printed electronic applications.

In this short communication, we aim to investigate the dispersion behavior of 2D Ge and Sn based (i.e. GeS and SnS) MMCs colloids produced via the LPE method. In particular, the dispersibility of GeS and SnS is investigated in nine different organic solvents and the dispersion yield and behavior of LPE inks is compared. It was revealed that the solvent surface energy and the Hansen solubility parameters of 2D MMCs played a crucial role towards the isolation of the dispersed inks. Our results elucidate the physiochemical interaction of 2D MMCs with various solvents during the LPE process, which could potentially affect a broad range of solution processable electronic applications of such materials.

2. **Results and discussion**

For the preparation of MMC colloids via LPE, powders of bulk germanium (II) sulfide (GeS) and tin (II) sulfide (SnS) were exfoliated and dispersed in various organic solvents, as detailed in the experimental section. In order to compare the exfoliation concentration/yield and the dispersion behavior in different solvents, the same amount of GeS and SnS powder was used. In particular, the exfoliated GeS and SnS nanosheets (3mg/ml) were tested in nine different solvents such as, NMP, IPA, DMSO, EtOH, DMF, MeOH, Acetone, THF, and DCM, most of which are commonly used for solution processable electronic and optoelectronic applications.

Among the different solvents studied, IPA and NMP resulted in the darkest dispersions for both GeS and SnS (**Figure 2**). In general, the optical appearance of the colloid was an indication of the suitability of the solvent in the isolation of exfoliated sheets. For this purpose, UV-Vis extinction spectroscopy was performed for each sample prepared. The extinction spectra of isolated MMCs show a broad light absorption profile (**Figure 1a,b**), including an extended absorption along the lower infrared region of the spectrum. For SnS in particular, the wide absorption range is accompanied with a shoulder at 420 nm (SnS). Such spectra are consistent with the reported results for both GeS and SnS dispersions [7, 22, 24], indicating the successful isolation of the MMCs sheets. On the other hand, the intensity of dispersed inks was varied among individual solvents. The highest value in extinction intensity complies with the high exfoliation rate or yield in the isolated ink. In our case, the highest absorption intensity was obtained in IPA for GeS and NMP in SnS sheets respectively, while the lowest in DCM, for both GeS and SnS ones.

The dispersibility of the MMCs ink in each solvent can be quantitatively estimated from the UV-vis extinction spectra. Based on the Beer-Lambert formula the absorbance A is,

$$A = \varepsilon L C \qquad (1)$$

Where, ε is the molar extinction coefficient, L is the path length and C is the concentration of the dispersed solution. In general, high C values indicate a good solvent and vice versa. Considering that the molar extinction coefficients is 2640 L g$^{-1}$ m$^{-1}$ (GeS ) [26] at 600 nm and 3.4 L g$^{-1}$ cm$^{-1}$ (SnS) [27] at 808 nm and the absorbances per unit length (A/L) can be extracted from the UV-vis spectra (**Table S1**), the respective C values of the isolated dispersions were estimated and are shown in **Table 1**.

**Table 1.** Physical and chemical properties [33, 34] of the various solvents used to produce MMCs inks in this study, together with the respective exfoliation concentrations.

| Solvents | Polarity | Boiling point (°C) | Dipole moment | Surface tension (mN/m) | Dielectric constant | Viscosity at 25°C (cP) | δ$_T$ (MPa$^{1/2}$) | GeS solubility (μg/mL) | SnS solubility (μg/mL) |
|---|---|---|---|---|---|---|---|---|---|
| **NMP** | Polar Aprotic | 202 | 3.75 | 40.1 | 32.2 | 1.67 | 22.9 | 7.03 | 23.08 |
| **IPA** | Polar Aprotic | 82 | 1.66 | 21.66 | 19.9 | 1.96 | 23.6 | 16.4 | 20.12 |
| **DMSO** | Polar Aprotic | 189 | 3.96 | 42.9 | 46.7 | 1.99 | 26.6 | 2.78 | 17.41 |
| **EtOH** | Polar Protic | 78 | 1.69 | 22.1 | 24.22 | 1.07 | 26.5 | 8.33 | 6.49 |
| **DMF** | Polar Aprotic | 153 | 3.82 | 37.1 | 32.70 | 0.79 | 24.9 | 2.20 | 6.83 |
| **MeOH** | Polar Protic | 65 | 1.70 | 22.7 | 32.70 | 0.54 | 29.6 | 8.12 | 6.86 |
| **Acetone** | Polar Aprotic | 56 | 2.88 | 25.2 | 20.70 | 0.31 | 19.9 | 2.87 | 0.40 |
| **THF** | Polar Aprotic | 66 | 1.75 | 26.4 | 7.58 | 0.48 | 19.4 | 2.18 | 11.2 |
| **DCM** | Polar Aprotic | 39.6 | 1.60 | 26.5 | 8.93 | 0.41 | 20.2 | 0.72 | 1.0 |

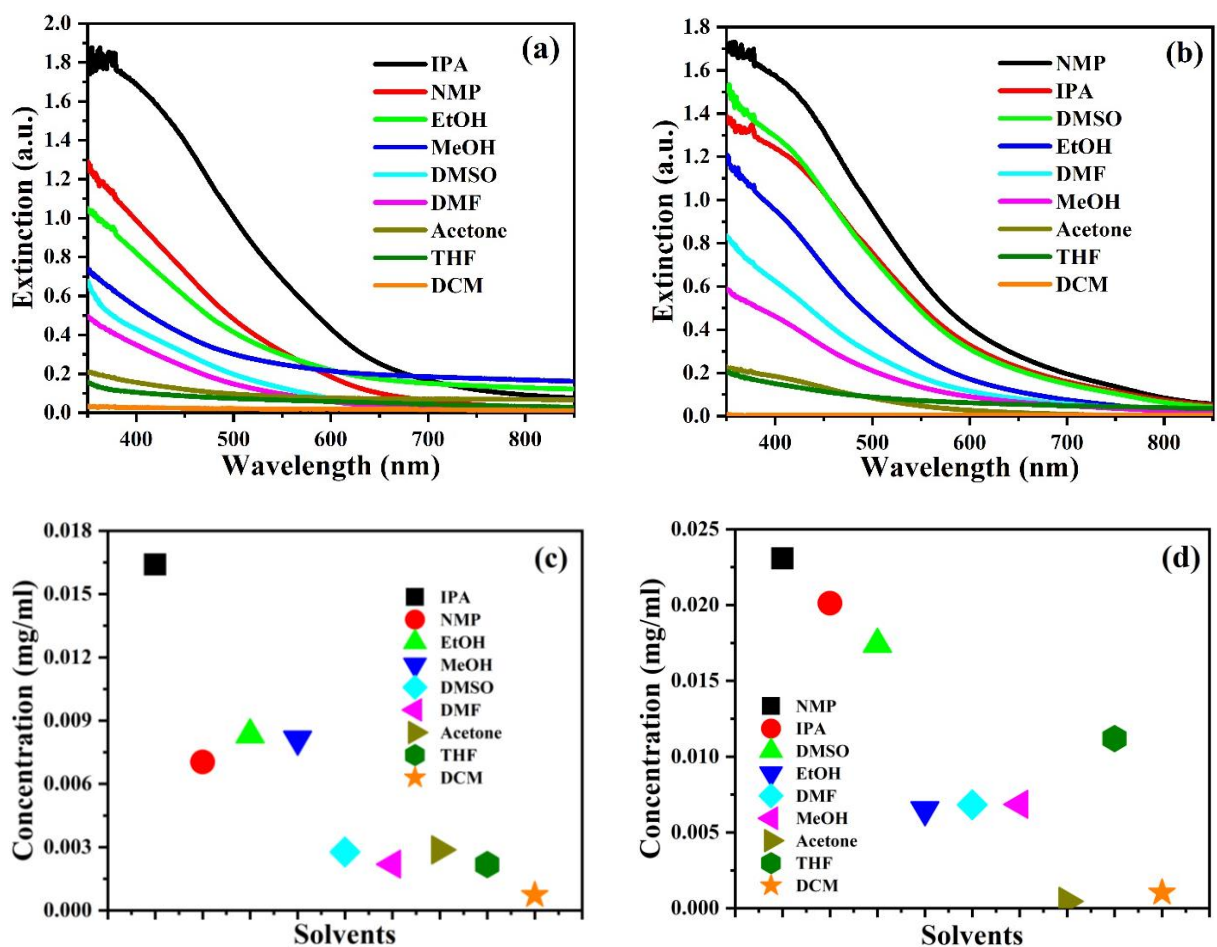

**Figure 1.** UV-Vis extinction spectra of liquid phase exfoliated GeS (a) and SnS (b) inks in nine different solvents. Calculated concentration, C, of GeS (c) and SnS (d) colloids in the nine different solvents tested.

The C values corresponding to each solvent are plotted in **Figure 1c** and **d**, showing that the highest exfoliated concentration was ~16.4 (in IPA) and ~23.08 μg/mL (in NMP) for GeS and SnS, respectively. Such values are comparable or even higher compared with those reported on other 2D materials (**Table 2**). In particular, for BP, which is an analogue to MMCs, the highest exfoliation yield attained to date was ~10 μg/mL in DMF and DMSO [28].

**Table 2.** Comparison of the conditions used and the corresponding exfoliation concentrations attained for the synthesis of GeS (this work), SnS (this work) and other 2D materials (literature) dispersions via LPE.

| Materials | Methodology | Exfoliation process | Tested solvents | Concentration (µg/ml) | Reference |
|---|---|---|---|---|---|
| **GO** | LPE | Bath sonication | 18 solvents | 8.7 (NMP) | [30] |
| **RGO** | LPE | Bath sonication | | 9.4 (NMP) | [30] |
| **MoS$_2$** | LPE | Sonication | EtOH/Water | 18 (EtOH/Water) | [31] |
| **WS$_2$** | LPE | Sonication | EtOH/Water | 32 (EtOH/Water) | [31] |
| **BP** | LPE | Cell Sonicator | DMF and DMSO | Up to 10 (DMF & DMSO) | [28] |
| **GeS** | LPE | Bath sonication | Hexene, ethanol, IPA, NMP, DMF, acetone, chloroform | - | [26] |
| | | Probe sonication | NMP | - | [24] |
| | | Bath sonication | NMP, IPA, DMSO, EtOH, DMF, MeOH, Acetone, THF, and DCM | 16.4 (IPA) 0.72 (DCM) | Present work |
| **SnS** | LPE | Bath sonication | NMP | - | [22] |
| | | Probe sonication and bath sonication | NMP | - | [32] |
| | | | IPA | - | [27] |
| | | Probe sonication | DMF | - | [23] |
| | | Bath sonication | Acetone | - | [7] |
| | | Bath sonication | NMP, IPA, DMSO, EtOH, DMF, MeOH, Acetone, THF, and DCM | 23.08 (NMP) 1.0 (DCM) | Present work |

The results presented in **Figure 1** comply with the optical appearance of post-centrifuged dispersion solutions, as shown by the digital photographs of **Figure 2a** and **c**, captured right after centrifugation. It can be observed that for both GeS and SnS the as-prepared colloids showed a fairly good dispersion quality in NMP, IPA, DMSO, EtOH, DMF, MeOH, Acetone, THF and DCM solvents. In order to verify the restacking and precipitation in post centrifuged

MMCs inks, pictures of the colloids were also taken after 96h of centrifugation (**Figure 2b, d**). It can be observed that all colloids, apart from that in EtOH, preserved their initial dispersion behavior, indicating that no significant nanosheets' restacking of takes place. Considering the solvent polarity, all nine solvents tested, including EtOH, exhibited high dipole moment values. This indicates that the solvent polarity is not the sole factor affecting the dispersibility and stability of MMCs inks in organic solvents [29]. Besides this, the variation in colloids' concentration among different solvents does not follow any specific trend with the solvent dipole moment (**Figure S1**).

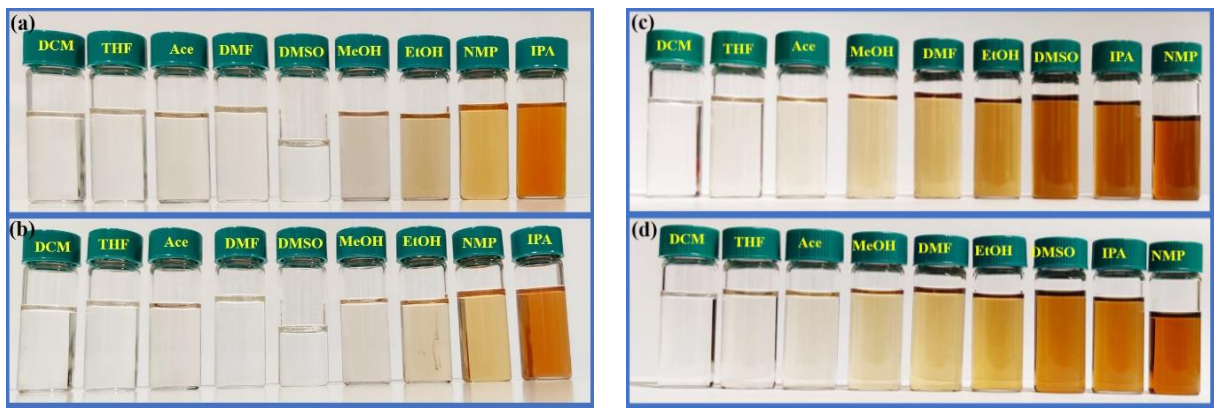

**Figure 2.** Digital pictures of isolated GeS and SnS dispersion inks in different solvents (a) Fresh GeS colloid just after centrifugation, (b) GeS dispersions after 96 h of centrifugation, (c) Fresh SnS colloid just after centrifugation and (d) SnS after 96 h of centrifugation.

Raman spectroscopy is a powerful tool to determine the successful isolation of ultrathin layers of 2D materials [35-37]. In particular, a blue shift in Raman spectra signify the successful isolation of thin 2D layers, while the Raman frequency difference between the in-plane and out-of-plane vibrational modes has been widely used to quantify the number of layers [4, 24, 38, 39]. MMCs belong to the orthorombic crystal structures, *Pnma* with $D_{2h}^{16}$ crystal symmetry [3, 38]; such crystals exhibit 24 phonon modes at the center of their Brillouin zone Γ and can be expressed as

$$\Gamma = 4A_g + 2B_{1g} + 4B_{2g} + 2B_{3g} + 2A_u + 4B_{1u} + 2B_{2u} + 4B_{3u} \tag{2}$$

where, $A_g$, $B_{1g}$, $B_{2g}$, and $B_{3g}$ are optically active Raman modes [7, 38, 40]. In particular, the $B_{3g}$ active mode corresponds to the in-plane vibration of adjacent layers parallel to the zig-zag direction, while the $A_g$ modes correspond to shared vibration of parallel layers in the armchair direction. **Figure 3** presents the respective Raman spectra of the isolated GeS and SnS nanosheets, where three prominent vibrational Raman modes are identified. It should be noted here that the bulk GeS exhibits three intense Raman peaks at ~209, ~236 and ~267 cm$^{-1}$, corresponding to the $B_{3g}$, $A_g^1$, and $A_g^2$ modes, respectively (**Figure S2**). On the contrary, all Raman vibrational modes in bulk SnS are silent and therefore no significant Raman peaks are present (**Figure S2**). On the other hand, the exfoliated GeS nanosheets exhibited three prominent Raman peaks appeared at ~213, ~240, and ~270 cm$^{-1}$, which are blue shifted by ~4 cm$^{-1}$ compared to the bulk counterpart. In general, the blue shift in Raman modes is ascribed to the reduction of dielectric screening effect, which is related to the thickness lowering in 2D materials [38, 40]. On the other hand, the inactive Raman modes of the bulk SnS appeared to be prominent in exfoliated SnS flakes (**Figure 3b**). In particular, the Raman spectra of isolated SnS exhibited three characteristic bands centred at ~156, ~187 cm$^{-1}$, and ~220 cm$^{-1}$ corresponding to $B_{3g}$, $A_g^1$, and $A_g^2$ vibrational modes, respectively. These results agree well with previously reported ones [7, 19], suggesting the successful isolation of thin layers of SnS.

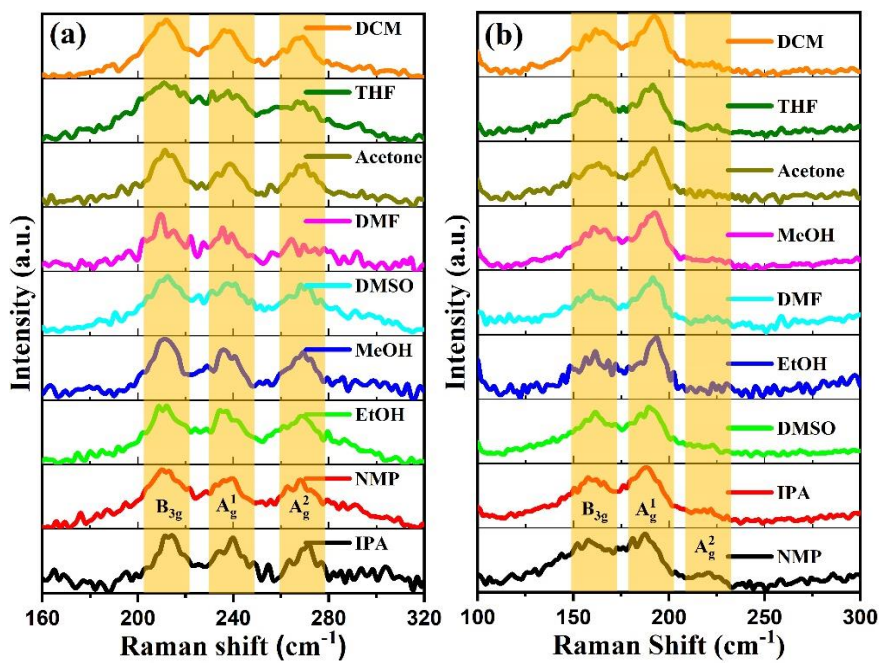

**Figure 3.** Raman spectra of isolated MMCs dispersion inks in various solvents (a) GeS and (b) SnS. The yellow solid bars represent the vibrational Raman modes present in isolated sheets.

AFM has been further employed to estimate the lateral and longitudinal dimensions of the isolated MMCs sheets obtained from the exfoliated GeS and SnS samples, in IPA and NMP solvents respectively. The corresponding AFM images (**Figure S3**) revealed an estimated average thickness of GeS nanosheets of ~3 nm in IPA and and ~13 nm in NMP, corresponding to ~5 and ~24 monolayers, respectively. On the other hand, the average thickness of SnS nanosheets were of ~4.5 nm in NMP and ~10 nm in IPA, corresponding to ~8 and ~18 monolayers respectively. In both cases the lateral dimensions of the isolated sheets were measured to be in the range from few tens to hundreds of nm.

It has been extensively demonstrated that the surface tension components of both the solvent and the nanosheets strongly affect the dispersion behavior and stability of LPE colloids [20, 21, 29]. For efficient exfoliation, the Gibbs free energy difference between the solvent and the 2D nanosheets should be as low as possible. Segev et al. [41] theoretically predicted the MMCs' surface energy and showed that the surface energy of GeS and SnS depends on crystal orientation. The corresponding minimum surface energy estimated to be 50 mN/m (0.05 J/m$^2$) and 80 mN/m (0.08 J/m$^2$) for GeS and SnS, respectively. Based on the above it may be postulated that the solvents exhibiting surface tension values that match those of dispersed nanosheets, are the most efficient ones. Indeed, it is shown that SnS nanosheets were efficiently dispersed in NMP, which, as shown in **Table 2**, exhibits a similar surface tension value. However, this is not the case for GeS nanosheets, which were shown to efficiently dispersed in IPA, although this solvent exhibits much lower surface tension than that predicted for GeS nanosheets (**Figure S2b**). In general, a large surface energy difference between the solvent and the 2D nanosheets could be an indication of the inefficiency of the exfoliation process in a specific solvent. However, there is currently not enough experimental evidence on the surface energy values for GeS and SnS nanosheets, which is a limiting factor against such a correlation.

The results presented here, demonstrate the efficient dispersion behavior of MMCs colloids in polar aprotic solvents such as NMP and IPA. The Hansen solubility parameters (HSP) of the solvents has been widely employed to describe the dispersion behavior of 2D materials [29]. In particular, the Hildebrand solubility parameter ($\delta_T$) is expressed as the sum of dispersion cohesion parameter ($\delta_D$), the polarity cohesion parameter ($\delta_P$) and the hydrogen bonding cohesion parameter ($\delta_H$) [42]

$$\delta_T^2 = \delta_D^2 + \delta_P^2 + \delta_H^2 \qquad (3)$$

In order to estimate the HSP values of GeS and SnS nanosheets, the following equation has been used

$$\delta_i = \frac{\sum_{solvent} C \delta_{i,solvent}}{\sum_{solvent} C} \qquad (4)$$

where, i is one of the D, P, H or T, while C is the GeS and SnS concentration (solubility) and $\delta_{i,solvent}$ is the i$^{th}$ HSP in a given solvent. The HSP values calculated for GeS are $\delta_D$ ~15.53

MPa$^{1/2}$, $\delta_P$ ~8.05 MPa$^{1/2}$ and $\delta_H$ ~16.8 MPa$^{1/2}$ for GeS, while, for SnS, $\delta_D$ ~18.11 MPa$^{1/2}$, $\delta_P$ ~14.74 MPa$^{1/2}$ and $\delta_H$ ~13.85 MPa$^{1/2}$. The calculated total solubility parameter $\delta_T$ using equation (3) yielded values of ~16.80 MPa$^{1/2}$ and ~19.31 MPa$^{1/2}$ for GeS and SnS, respectively. Similar values of the HSP of solvent and solute, is an important criterion towards the identification of an efficient solvent. This may explain the highest solubility obtained in IPA (GeS) and NMP (SnS), which have similar $\delta_T$ and solubility parameters (Table 1).

In order to examine the long-term stability of the MMCs colloids, the as-exfoliated dispersions were stored for more than three months under ambient conditions. **Figure 4** shows the digital images of the GeS and SnS colloids stored more than three months under ambient conditions. It can be clearly observed that the GeS and SnS colloids have retained its excellent dispersibility in IPA and NMP solvents respectively. This is in contrast to the other solvents cases, which have shown a moderate but visible precipitation at the bottom of glass vials stored. The precipitation observed can be primarily attributed to the restacking of the layered flakes [30, 43]. It is worth mentioning that the SnS dispersion in NMP is significantly more stable than that of GeS, probably due to electrostatic repulsion promoted by its positive surface energy. Furthermore, the GeS and SnS have shown a great merit of exfoliation yield and stability in polar aprotic solvents, which can be attributed to the electrostatic repulsion due to surface charges. This is further supported by our general observation that the nonpolar solvents, such as chlorobenzene or dichlorobenzene, are inappropriate for MMCs exfoliation via LPE.

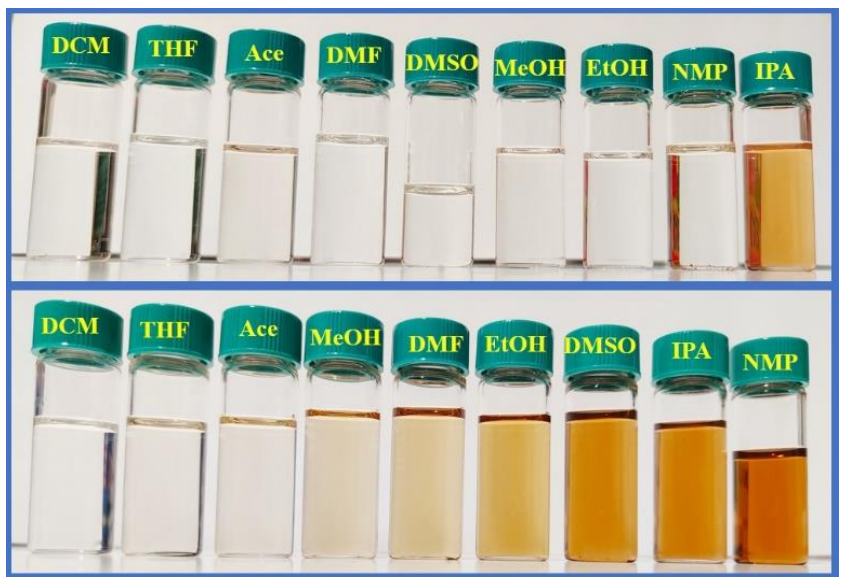

**Figure 4.** Digital pictures of isolated GeS (top panel) and SnS (bottom panel) dispersions following more than three months of storage in ambient conditions. The long-term stability of the inks in different solvents is evident.

3. **Conclusions**

In summary, we have performed liquid phase exfoliation of GeS and SnS in in nine different organic solvents and the dispersion behavior of colloids obtained has been evaluated. It is observed that the isolated dispersed flakes are few layers thin with lateral dimensions of a few hundreds of nanometers. Among the organic solvents tested, IPA was found to be the best for SnS, and NMP for GeS. In particular, the obtained GeS and SnS dispersions in IPA and NMP respectively showed the highest concentrations of ~16.4 μg/mL and ~23.08 μg/mL, respectively. Besides this, it is shown that the isolated dispersions are highly stable even after months of storage in ambient conditions. Our results shed light on the nature of dispersion behavior and stability of MMCs inks, indicating the great potential of such materials to be used for future solution processable electronic applications.

**Methods.**

*Liquid phase exfoliation of MMCs.*

Tin (II) sulfide (>99.99%) and germanium (II) sulfide (99.99%) granular trace metals basis was purchased from Sigma Aldrich, USA and used without further purification. N-Methyl-2-pyrrolidone (NMP, 99.5%), 2-propanol (IPA, ≥99.9%), Dimethyl Sulfoxide (DMSO, ≥99.7%), absolute Ethanol (EtOH), Dimethylformamide (DMF, ≥99.8%), Methanol (MeOH, (>99.9%), Acetone, Tetrahydrofuran (THF, ≥99.9%), and Dichloromethane (DCM, ≥99.9%) were received from Sigma Aldrich, Honeywell and Fisher scientific, and used without further purification for exfoliation. In order to produce the MMCs ink, the bulk powder of SnS and GeS were dissolved and dispersed in various organic solvents. In brief, bulk powder of GeS and SnS (3 mg/ml) was dissolved in nine different solvents in nitrogen flushed glass vials and sealed with nitrogen to avoid any atmospheric impact during LPE. The sealed solution vials of SnS and GeS were ultrasonicated in a Elma S 30 H bath sonicator (Elma Schmidbauer GmbH, Germany) under 80 W power and 37 kHz frequency for 20h and 5h, respectively. The dispersed dark brown solutions were centrifuge to remove the bulk component in atmosphere. Specifically, SnS and GeS solutions were centrifuged at 8000 and 6000 rpm for 15 min, respectively. The centrifuged suspensions were collected as a ink and used for further investigations.

*Extinction spectroscopy*

Extinction spectra of the isolated MMCs dispersions were recorded with a Perkin Elmer, Lamda 950 UV/VIS/NIR spectrometer, USA. The path length of the cuvette was 10 nm.

*Raman spectroscopy*

The Raman spectroscopic analysis of an isolated MMCs were performed with 473 nm laser excitation (Thermo Scientific) in the back-scattering geometry at ambient conditions at 300 K. The spot size of the laser was ~1 μm. The incident beam was focused on the sample using a 50X long-working distance objective. The system was calibrated with a Si substrate peak at 520 cm$^{-1}$. All the samples were prepared by drop casting on cleaned SiO$_2$ substrate.

*Atomic force microscope*

Tapping mode AFM images were captured in an instrument of Multimode Atomic Force Microscope from Digital Instruments, Bruker, USA. The samples were prepared by drop casting dispersion of layered MMCs inks on cleaned Si substrate. The casted samples were dried/annealed and placed on the microscope stage to scan.

**Acknowledgements**

This work was supported by the EC HORIZON 2020 projects 'NFFA-Europe', under Grant agreement No. 654360 and 'MouldTex', under Grant Agreement No. 768705.

**Appendix A. Supplementary material**

Supplementary data associated with this article can be found, in the online version at the journal website.

**Graphical abstract**

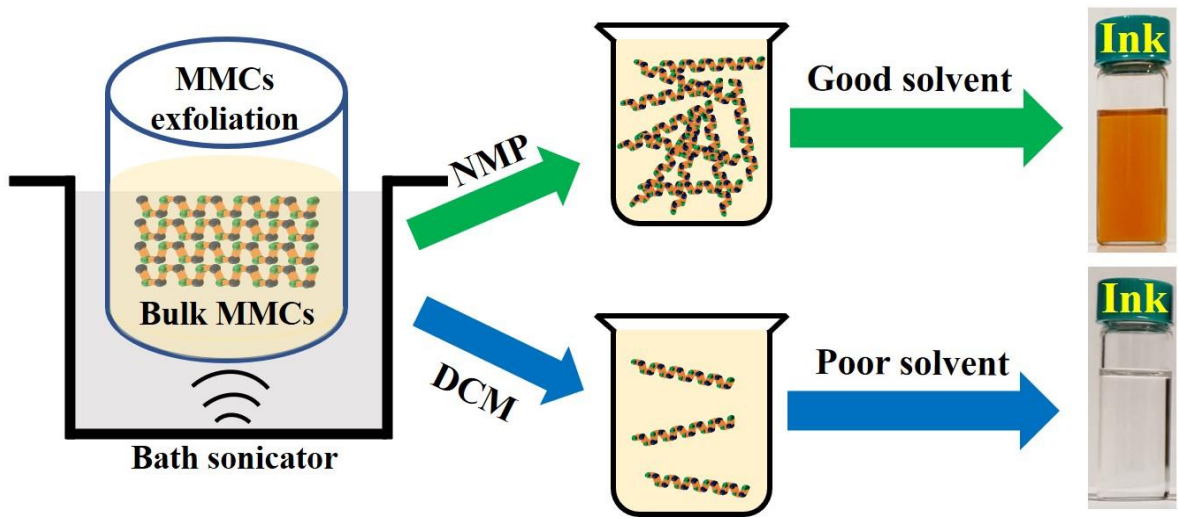

**Supporting information……..**

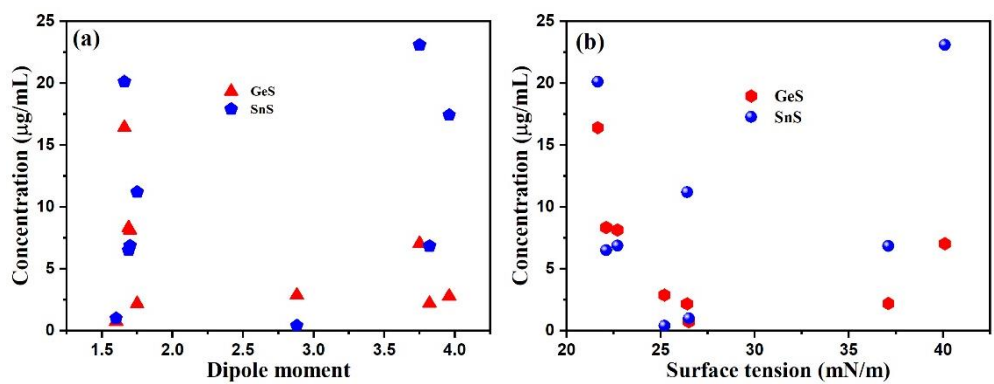

**Fig. S1.** Concentration variation in isolated GeS and SnS inks exfoliated in various organic solvents with (a) dipole moment and (b) surface tension.

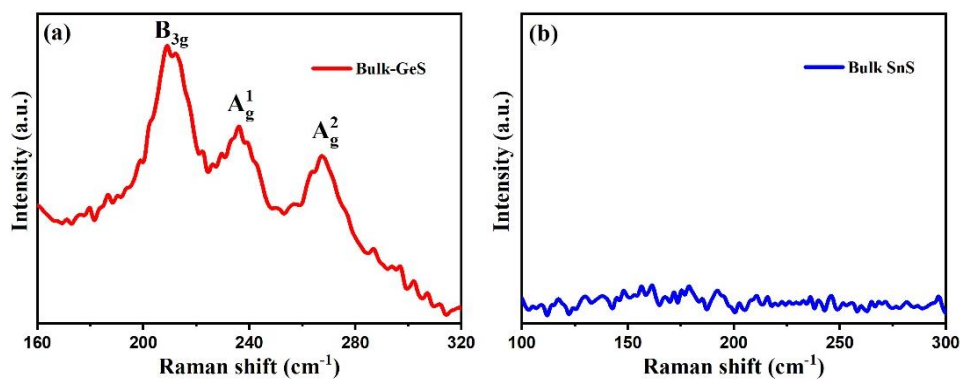

**Fig. S2.** Raman spectra of bulk (a) GeS and (b) SnS powder used for exfoliation. The spectra were collected with an excitation of 473 nm laser line.

**Table S1.** Optical absorbance per cell length, A/l (a.u.), for GeS and SnS in various organic solvents

|   | GeS | | SnS | |
|---|---|---|---|---|
|   | Solvent | A/l at 600 nm a.u. | Solvent | A/l at 808 nm a.u. |
| 1 | IPA | 0.0432 | NMP | 0.00785 |
| 2 | NMP | 0.0185 | IPA | 0.00684 |
| 3 | EtOH | 0.0219 | DMSO | 0.00592 |
| 4 | MeOH | 0.0214 | EtOH | 0.00221 |
| 5 | DMSO | 0.0073 | DMF | 0.00232 |
| 6 | DMF | 0.0058 | MeOH | 0.00233 |
| 7 | Ace | 0.0075 | Ace | 0.00014 |
| 8 | THF | 0.0057 | THF | 0.00381 |
| 9 | DCM | 0.0019 | DCM | 0.00034 |

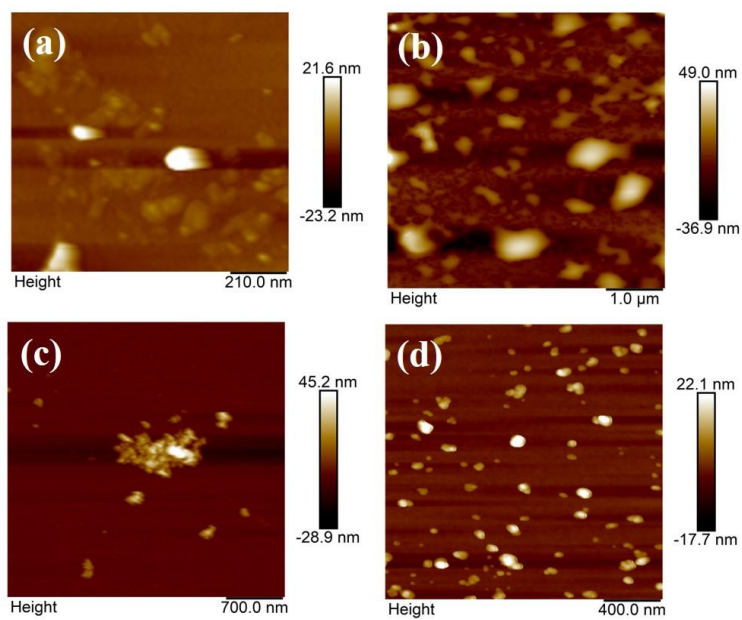

**Fig. S3.** AFM images (a, b) GeS in IPA and NMP and